\documentclass[10pt]{article}
\usepackage[comma]{natbib}
\usepackage[dvips]{graphicx}
\newcommand{\sradi}{R_{\odot}}
\title{Exploring a deep meridional flow hypothesis for a circulation
  dominated solar dynamo model}
\author{G. A. Guerrero\footnote{email:guerrero@astro.iag.usp.br},
J. D. Mu\~{n}oz\footnote{djmunozc@unal.edu.co},  Elisabete M. de
Gouveia Dal Pino\footnote{email:dalpino@astro.iag.usp.br}}  
\begin{document}

\maketitle

\abstract{Circulation-dominated solar dynamo models, which employ a
helioseismic rotation profile and a fixed meridional flow, give a
good approximation to the large scale solar magnetic phenomena,
such as the 11-year cycle or the so called Hale's law of
polarities. Nevertheless, the larger amplitude of the radial shear
$\partial \Omega / \partial r$ at the high latitudes makes the
dynamo  to produce a strong toroidal magnetic field at high
latitudes, in contradiction with  the observations of the sunspots
 (Sporer`s Law).  A possible solution was proposed by \cite{nandy2002}
 in which a deep meridional flow can 
conduct the magnetic field inside of a stable layer (the radiative
core) and then allow that it erupts just at lower latitudes.
Although they obtain good results, this hypothesis generates new
problems like the mixture of elements in the radiative core (that
alters the abundance of  the elements) and the transfer of angular
momentum.  We have recently explored this hypothesis in a
different approximation, using the magnetic buoyancy mechanism
proposed by \cite{dikchar99} and
found that a deep meridional flow pushes the maximum  of the
toroidal magnetic field towards the solar equator, but, in
contrast to \cite{nandy2002}, a second zone of maximum fields remains
at the poles \citep{guerrero04}. In that work, we have also introduced
a  bipolytropic density profile in order to better reproduce the
stratification in the radiative zone. We here review these results and
also discuss a new possible scenario where the tachocline has an
ellipsoidal shape, following early helioseismologic observations
\citep{char99}, and find that the modification of the geometry of the
tachocline can lead to results which are in good agreement with
observations and opens the possibility to explore in more detail,
through the dynamo model, the place where the magnetic field could be
really stored.}

\maketitle


\section{Introduction}

Over the last few years, large scale solar magnetic field
features, such as the 11 years sunspots cycle or polarity
inversions of the magnetic field, have been successfully explained
by solar dynamo models that include a solar like differential
rotation and meridional circulation as the main ingredients. Since
helioseismic measurements have indicated  the presence of a
tachocline, it has been a common believe that the dynamo action
takes place in this thin layer with substantial radial shear. In
the tachocline, a poloidal magnetic field is stretched by the
solar rotation such that a belt of strong toroidal magnetic field
is formed around the solar equator, theoretical arguments suggest
that this field is not homogeneous but  mainly concentred in
magnetic flux tubes surrounded for less magnetized plasma. The
strong magnetic pressure can make these flux tubes rise to the
surface by the action of magnetic buoyancy. There, the poloidal
field is possibly regenerated by the decay of the bipolar magnetic
regions. This poloidal field is then advected by meridional
circulation towards the solar poles and then to the deep layers,
where the full cycle will be completed. It is not completely
understood yet what is the profile of the meridional circulation,
but its crucial role on the dynamo process has been recently
recognized and the models that include it have been named
circulation dominated solar dynamo (CDSD) models.

Most of the  models that consider  a differential rotation profile
derived from helioseismic inversions, present an interesting
problem, the radial shear $\partial \Omega / \partial r$ (where
$\Omega$ is the angular velocity and $r$ is the radial distance)
is larger at higher latitudes than at lower ones so that a strong
toroidal field is expected to be generated in regions closer to
the poles and therefore sunspots should appear at higher
latitudes, contradicting the observations (Sporer`s Law).
Apparently this was an unavoidable problem and no change in the
parameters could solve it \citep{dikchar99,kuketal01} until
\cite{nandy2002} proposed a new possible 
scenario in  order to reduce the amplitude of the toroidal field
at higher latitudes. They suggested that the meridional flow could
go a little deeper than in the previous models. With a flow going
below the tachocline, the toroidal magnetic field is stored in a
more stable layer and advected by meridional circulation to lower
latitudes from where it emerges to the tachocline and the
convective zones and undergoes magnetic buoyancy.
This model (see also \cite{chatetal04}) results a magnetic field
distribution which is in qualitative agreement with the
observations of the sunspots distribution, though it may affect
the mixture of the elements and the observed abundances, and also
the angular momentum transfer, but these specific potential
problems will not be addressed in the present work.

There is also another fundamental difference in the way by which
the model of \cite{nandy2002} and the previous ones
treat the buoyant process. One of the aims in these models is to
produce a flux tube of strong toroidal field ($B_c=10^5$ G) in a
subadiabatic layer that becomes buoyant unstable and emerges to
the surface in a time which is much smaller (a month) than the
characteristic time of one magnetic cycle (of the order of ten
years). While \cite{dikchar99} used a
nonlocal treatment, making the poloidal field to regenerate closer
to the surface proportionally to the toroidal field at the base of
the solar convection zone, \cite{nandy2002}
employed a numerical procedure in which whenever the toroidal
magnetic exceeded $B_c$ at the base of the solar convection zone, a
fraction $f$ of it is made artificially to erupt to the surface at time
intervals $\tau=8.8 \times 10^5$ \citep{nandy2001}. In a recent work 
\cite{guerrero04}, we have developed a CDSD model using a solar
differential rotation like \cite{nandy2002}, but
employed the magnetic buoyancy mechanism used by \cite{dikchar99}. We
have also considered a more realistic bipolytropic 
density distribution \citep{gpinzon} in order to better reproduce the
subadiabatic stratification of the radiative zone. 
 We have found that properties such as the 11 years cycle,  the
 inversion of polarity, and the equatorward migration of branches of
 strong toroidal field and the poleward migration of a weak poloidal 
field, are correctly reproduced, however even using a deep
meridional flow, a strong toroidal field still remains in the
poles.
 The
introduction of the bipolytropic profile has not modified either
the morphology of the butterfly diagrams or the branches of strong
toroidal field  in the poles.

In the  present work, we have tried to improve our model by
modifying the geometry of the tachocline taking into account the
early helioseismic results of \cite{char99}
according to which the tachocline may have a prolate
shape. In section \S\ref{math}, we present the mathematical
formalism of our model. The numerical details can be found in
\citep{guerrero04}. Section \S\ref{results} contains the main
results of our CDSD model with a prolate  tachocline and, finally,
in \S\ref{conclusions}, we draw our conclusions.

\section{The Model}\label{math}

Our code solves the MHD induction equation which governs the evolution
of the solar dynamo:
\begin{equation}
\frac{\partial \bf{B}}{\partial t}=\nabla \times (\bf{U}\times
\bf{B})+\eta \nabla^2\bf{B}\label{eq1} \quad.
\end{equation}
By assuming spherical symmetry, the magnetic and velocity fields can
be writen as
\begin{eqnarray}
{\bf B}&=&B(r,\theta,t) + \nabla \times
(A(r,\theta,t))\label{eq2},\\
{\bf U}&=&{\bf u}(r,\theta)+r\sin \theta \Omega (r,\theta)
\label{eq3} \quad,
\end{eqnarray}
where $B(r,\theta,t)$ and $\nabla \times (A(r,\theta,t))$
correspond to the toroidal and the poloidal components of the
magnetic field, respectively; $\Omega$ is the angular velocity,
$\bf{u}=u_r+u_{\theta}$ is the velocity in the meridional plane,
and $\eta$ is the magnetic diffusivity. Replacing equations
(\ref{eq2}) and (\ref{eq3}) in the induction equation (\ref{eq1})
and separating the poloidal and toroidal components of the
magnetic field, we obtain
\begin{eqnarray}
\frac{\partial A}{\partial t}+\frac{1}{s}({\bf u}
\cdot\nabla)(sA)=\eta_p(\nabla^2-\frac{1}{s^2})A +
S_1(r,\theta,t)\label{eq4} \quad ,\\
\frac{\partial B}{\partial t}+\frac{1}{r}[\frac{\partial}{\partial
    r}(ru_rB)+\frac{\partial}{\partial
    \theta}(u_{\theta}B)]\label{eq5}=({\bf B_p} \cdot
\nabla) \Omega\\\nonumber
-\nabla \eta_t \times \nabla \times
B+\eta_t(\nabla^2-\frac{1}{s^2})B\nonumber \quad,
\end{eqnarray}
where $s$$=$$r\sin\theta$ and $\bf{B_p}$$=$$\nabla \times A$. The
values and profiles of $\eta_p$ and $\eta_t$ will be disscused in the
section \ref{magdiff}.  In the right side of eq. (\ref{eq4}), 
$S_1(r,\theta,B_{\phi})$ is a source term (see below).  

\subsection{Differential rotation}
The solar differential rotation inferred from  helioseismology can
be described by the following equation
\begin{equation}
\Omega(r,\theta)=\Omega_c+\frac{1}{2}[1+erf(2\frac{r-r_c}{d_1})](\Omega_s(\theta)-\Omega_c)
\quad .
\label{eq6}
\end{equation}
Here,
$\Omega_s(\theta)$$=$$\Omega_{Eq}+a_2\cos^2\theta+a_4\cos^4\theta$
is the latitudinal differential rotation at the surface and
$erf(x)$ is an error function that confines the radial shear to a
tachocline of thickness $d_1$$=$$0.05 \sradi$. In this expression,
a rigid core rotates uniformly with angular velocity
$\Omega_c/2\pi$$=$$432.8$. The other values are
$\Omega_{Eq}/2\pi$$=$$460.7$, $a_2/2\pi$$=$$-62.9$,
$a_4/2\pi$$=$$-67.13$ nHz, and $r_c$$=$$0.7\sradi$.

\subsection{Meridional circulation}
 As in \citep{dikpati94,choudhuri95,nandy2002},
 we assume a single convection cell for each meridional quadrant
\begin{equation}
\rho(r) \bf{u}=\nabla \times [\psi(r,\theta)e_{\phi}]\label{eq7} \quad,
\end{equation}
where $\psi$ is the stream function given by
\begin{eqnarray}
\psi r\sin\theta &=& (r-R_b)\psi_0
\sin[\frac{\pi(r-R_b)}{(\sradi-Rb)}]\\\nonumber
&\times&(1-e^{-\beta_1r\theta^{\epsilon}})(1-e^{\beta_2r(\theta-\pi/2)})\\\nonumber
&\times&e^{[(r-ro)/\Gamma]^2}\label{eq8} \quad,
\end{eqnarray}
and $\rho$ is the density profile for an adiabatic gaseous sphere with a
specific heat ratio coefficient $\gamma$$=$$5/3$ ($m$$=$$1.5$). Then
\begin{equation}
\rho(r)=C(\frac{\sradi}{r}-0.95)^m\label{eq9} \quad ,
\end{equation}
where we chose $C$$=$$3.60 \times 10^{-3}$ g cm$^{-3}$ as the
surface density value \citep{gpinzon}. The coefficient $\psi_0$
was chosen in such a way that the maximum latitudinal velocity at
middle latitudes is $20$ m s$^{-1}$. The other values in  eq.
(\ref{eq8}) are: $\beta_1$$=$$1.65 \times 10^{10}$ cm$^{-1}$,
$\beta_2$$=$$2.2\times10^{10}$ cm$^{-1}$,
$\epsilon$$=$$2.0000001$, $r_o$$=$$(\sradi-R_{min})/4.15$, and
$\Gamma$$=$$3.47\times 10^{10}$ cm. Here, $R_{min}$ is the minimum
$r$ coordinate value for the integration range, and the free
parameter $R_b$ is the maximum depth of the return flow (see
\citep{dikpati94} for more details).

\subsection{Magnetic buoyancy (MB) and the alpha effect}

As suggested in previous simulations of magnetic flux tubes
\citep{dsilva93,fan93,caligari95,calig98}, a tube of strong
magnetic field surrounded by a diffuse field can become buoyant
unstable and  arise to the surface where it is twisted by the
coriolis force. It has been shown that a $10^5$ G magnetic field
becomes unstable and emerges to the surface in time intervals of
about one
 month to form bipolar regions with the appropriate  tilt angles.
We introduce this effect in a simplified way by using a non-local
regeneration of the toroidal magnetic field followed by an alpha
effect which is concentrated close to the surface, so that:
\begin{eqnarray}
  S_1(r,\theta;B)&=&\frac{S_o}{4}
  B(r_c,\theta,t)[1+erf(\frac{r-r_2}{d_2})]\label{eq10}\\\nonumber&\times&[1-erf(\frac{r-r_3}{d_3})]\\\nonumber&\times&[1+(\frac{B(r_c,\theta,t)}{B_0})^2]^{-1}
  \cos\theta \sin\theta \quad.
\end{eqnarray}
 Where the parameters are $r_2$$=$$0.95$$\sradi$, $r_3$$=$$\sradi$,
and $d_2$$=$$d_3$$=$$0.025$$\sradi$.

\subsection{Magnetic diffusivity}\label{magdiff}

The magnetic diffusivity is larger in  regions of stronger
magnetic field. Therefore,  we assume a larger diffusivity for the
poloidal component, as suggested  by \cite{chatetal04}:
\begin{equation}
\eta_p(r)=\eta_c + \frac{\eta_{cz}}{2}[1+erf(2\frac{r-r_c}{d_1})] \quad,
\label{eq11}
\end{equation}
with  $\eta_c$$=$$2.2\times 10^9$ and $\eta_{cz}$$=$$0.2\times
10^{11}$ cm$^2$ s$^{-1}$, and $r_c$$=$$0.7$$\sradi$, and smaller
diffusivity for the toroidal component:
\begin{equation}
\eta_t(r)=\eta_c + \frac{\eta_{cz1}}{2}[1+erf(\frac{r-r_{c1}}{d_1})]+\frac{\eta_{cz}}{2}[1+erf(\frac{r-r_{c2}}{d_1})]\quad,\label{eq12}
\end{equation}
where $\eta_{cz1}$$=$$4\times 10^{10}$ cm$^2$s$^{-1}$,
$r_{c1}$$=$$0.72$$\sradi$, and $r_{c2}$$=$$0.95$$\sradi$.

\section{Results}\label{results}

We have solved the  equations  above by means of the ADI method in
a two-dimensional mesh of $128$$\times$$128$ spatial divisions,
with $0.55 \sradi \le r \le 1 \sradi$, and $0 \le \theta \le
\pi/2$. In the model of Fig. 1, the meridional flow is allowed to
penetrate down to  the base of the solar convection zone
($r$$=$$0.675\sradi$).

\begin{figure}[ht]
\begin{centering}
  \includegraphics[height=.2\textheight]{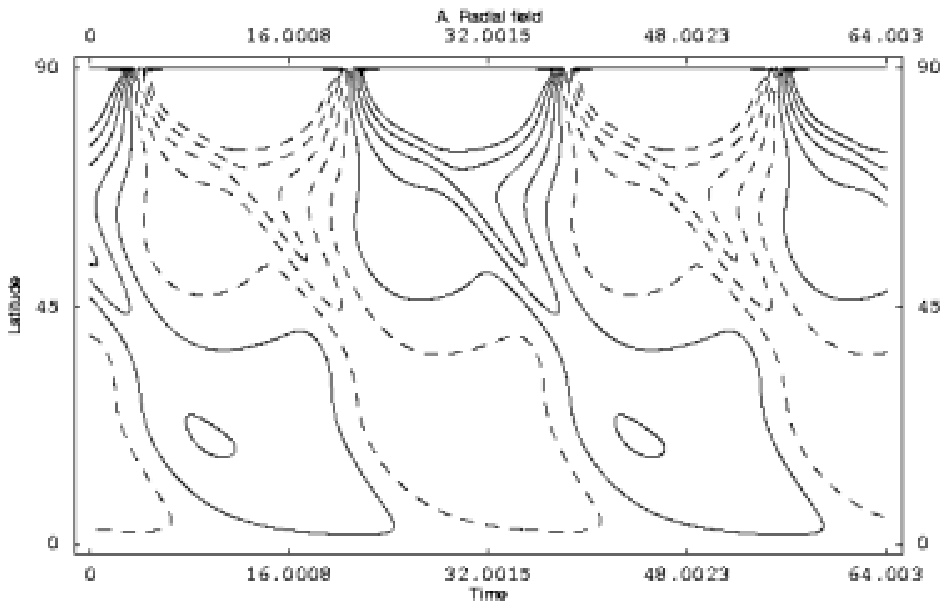}
  \includegraphics[height=.2\textheight]{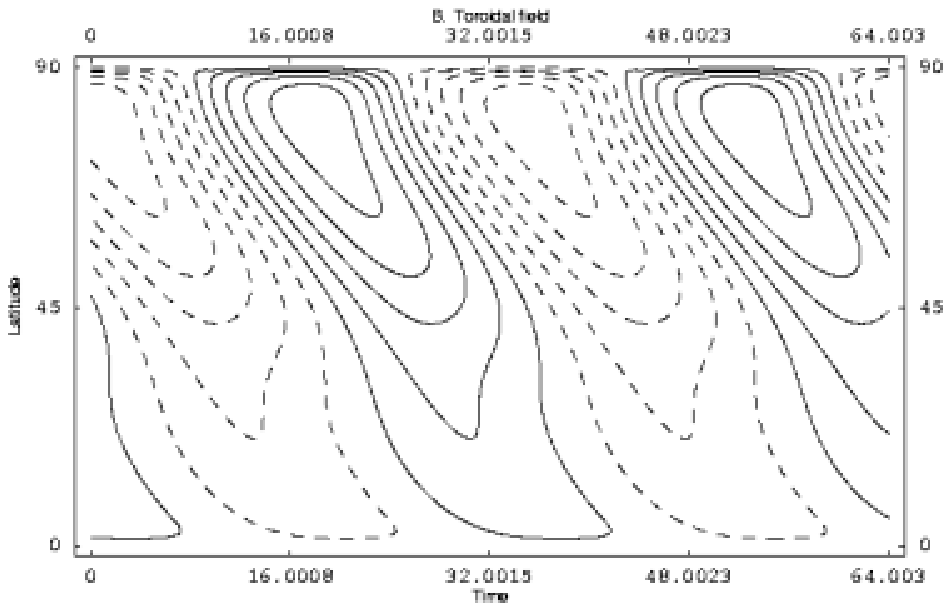}
  \caption{ Butterfly diagrams for (A)  the
  radial field at the surface ($r$$=$$\sradi$) and (B) the toroidal
  magnetic field at the base of the solar convection zone
  ($r$$=$$0.7\sradi$). The contours are
  equally  spaced, with solid (dashed) lines for positive (negative)
  values. Time is in years and the latitude is in degrees. The maximum toroidal
  magnetic field at the base of the SCZ is located at latitudes
  between $75^\circ$  and $90^\circ$ with weak branches
  migrating toward the equator. With this result and based on
  magnetic flux tubes simulations, sunspots will appear mostly in regions
  close to the poles.}\label{fig1}
\end{centering}
\end{figure}

\subsection{Deep Meridional Flow}
If we allow the flow to go  deeper ($r$$=$$0.61\sradi$), a better
distribution of the toroidal field lines is obtained, however, a
polar branch also appears, in contrast to the observations and to
the results \cite{nandy2002} (Fig. 2).

\begin{figure}[ht]
\begin{centering}
  \includegraphics[height=.2\textheight]{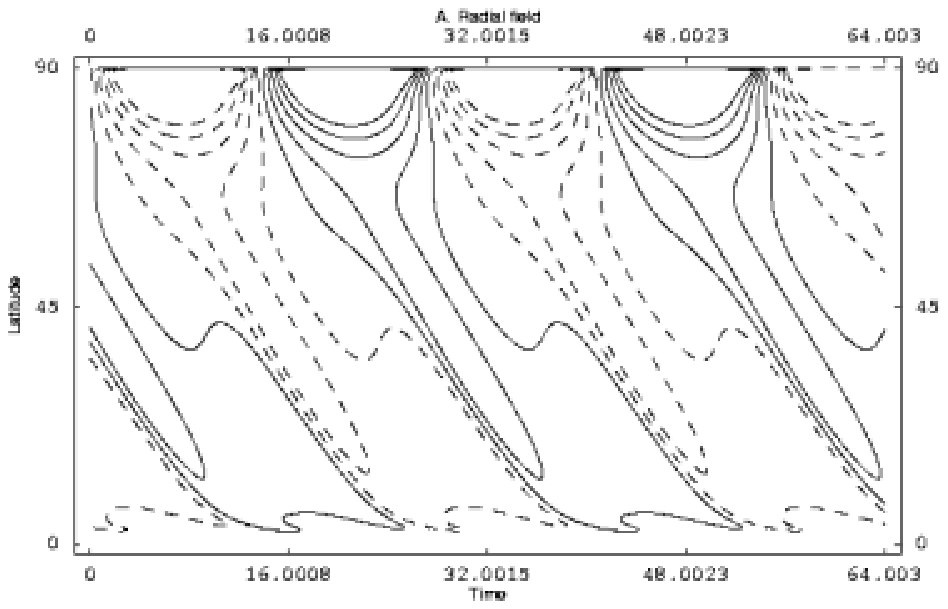}
  \includegraphics[height=.2\textheight]{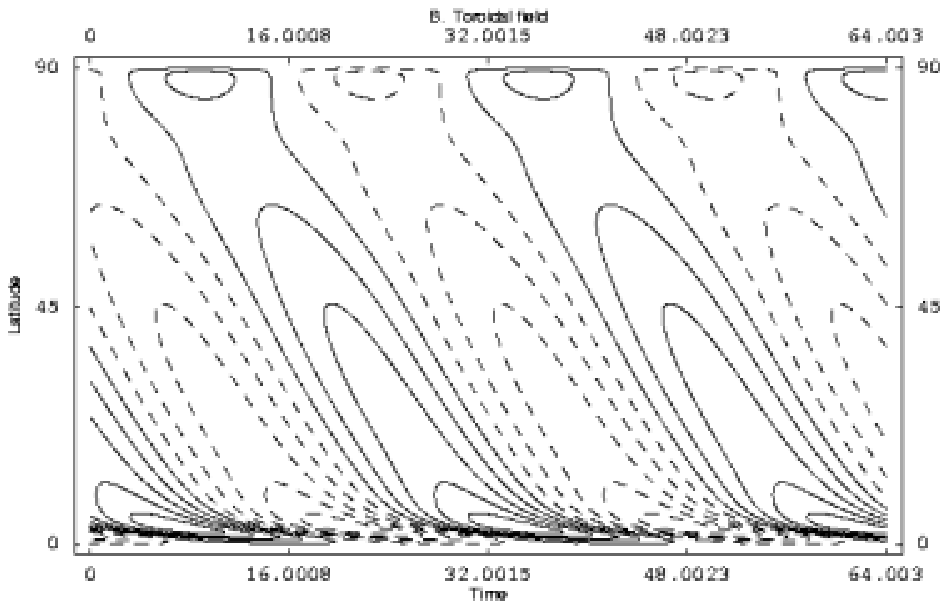}
  \caption{ The same as in Fig. 1, but for a deeper meridional flow.
  If most of the magnetic field is dragged below the tachocline, only
  a small portion emerges to the surface by the action of the magnetic
  buoyancy force.}\label{fig2}
\end{centering}
\end{figure}

\subsection{Exploring an ellipsoidal shape of the tachocline}

\begin{figure}[ht]
\begin{centering}
  \includegraphics[height=.2\textheight]{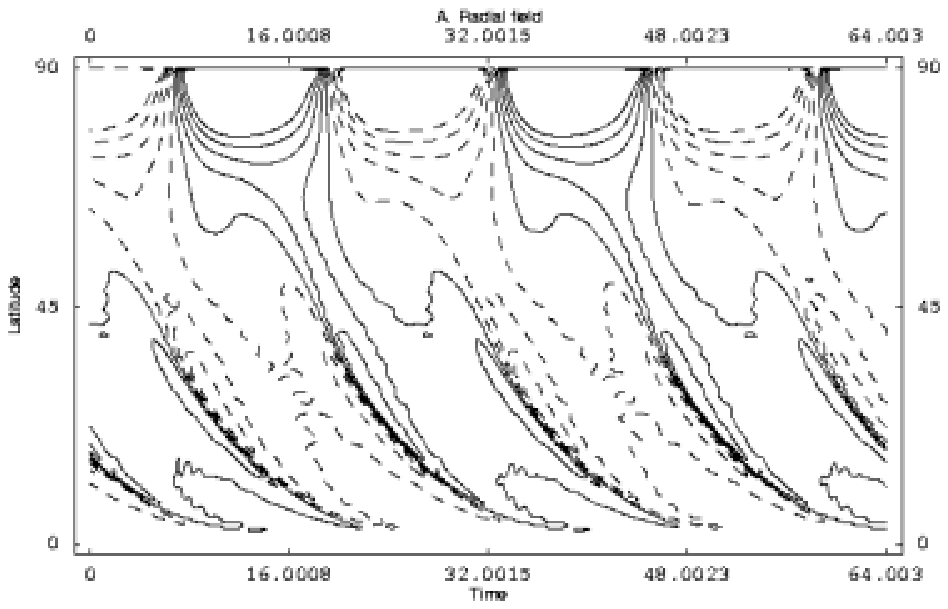}
  \includegraphics[height=.2\textheight]{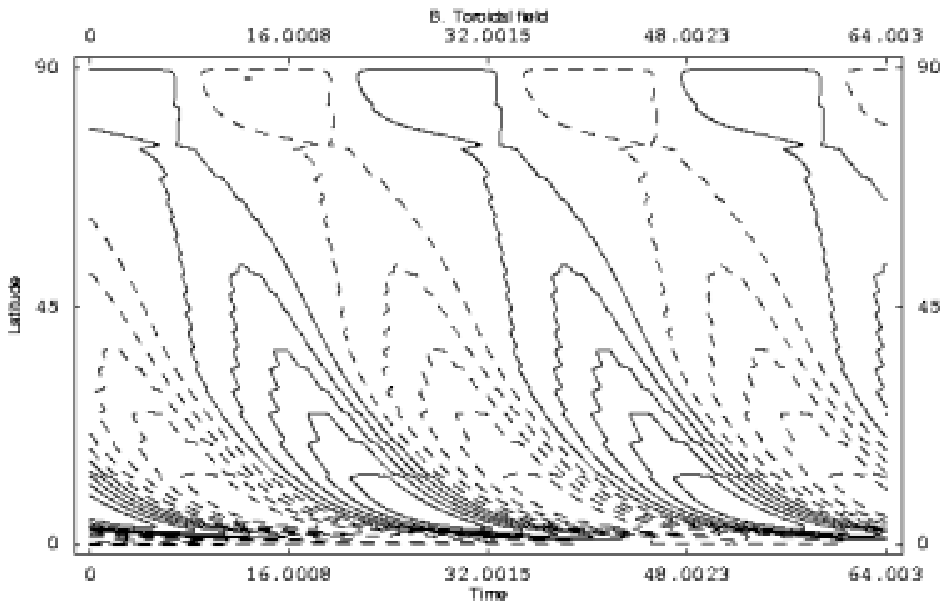}
  \caption{ Butterfly diagrams for (A) the radial and (B) the toroidal
  magnetic fields assuming a prolate  tachocline and a source term
  proportional to the toroidal magnetic field just above the
  tachocline.}\label{fig3}
\end{centering}
\end{figure}

Another way to test the deep meridional flow hypothesis is to
change the form of the tachocline. In the helioseismic results of
\cite{char99}, they have reported a possible prolate shape of the
tachocline. Though the full properties of the tachocline still remain
unknown, we may here modify  the shape of this layer by replacing the
constant term $r_c$ in the equations above for a term that depends on
$r$ and $\theta$ $r_c(r,\theta)$: 

\begin{equation}
r_c(r,\theta)=\frac{r_{cx} r_{cy}}{\sqrt{r_{cx}^2\cos^2\theta +
r_{cy}^2\sin^2\theta }} \label{eq13},
\end{equation}

where $r_{cx}$ and $r_{cy}$ are the semi-axis of an ellipse. 
If we first vary  the term $r_c$ only in the
differential rotation eqs. (6) and (10), in order to assure a
substantial vertical shear, but maintaining it constant in eqs. (11)
and (12), this configuration produces a prolate  tachocline  
($r_{cx}$$=$$0.67\sradi$ and $r_{cy}$$=$$0.73\sradi$) that generates  
weak polar branches (with magnetic fields of the order of  $10^4$ G)
and strong equatorward branches of toroidal field, in good agreement
with the observations (see Fig. \ref{fig3}). In this case, the
poloidal magnetic field, which is advected by meridional flows, goes
deeper at the high latitudes within the more stable radiative layer.
The toroidal field will diffuse more rapidly due to
the longer values of the coefficient of diffusion in these latitudes.
Advection then makes the rest of the process, draging this field to
reach the tachocline and the convection zone, and arise to the
surface. This scenario at which the 
flux goes deeper is more reliable because the mixture of elements 
occurs in a thinner layer within the radiative zone. 
If we now allow the $r_c$ term to vary according to eq. (13) also
in the equations (11) and (12) and test this prolate configuration,
we recover the results of the previous section (Fig. 2). A more detailed
analysis will be presented elsewhere in a forthcoming paper.

\section{Conclusions}\label{conclusions}

We have recovered the \cite{nandy2002} results
and qualitatively reproduced the observed latitudinal distribution
of the magnetic fields in the Sun  assuming a different source
formulation \citep{dikchar99}.  The most important results here
found may be summarized as follows:
\begin{enumerate}
\item If the meridional flow is confined to the convective zone ($r$$>
  $$0.675\sradi$), the  sunspots are mainly concentrated near
  the poles.
\item If the flow is allowed to penetrate deeper down to
$0.61\sradi$, the model gives
  a better result since a branch of  maximum toroidal field is
  produced at  low latitudes in agreement with the
  observations. However, in this case, another branch also appears
  near the poles which is inconsistent with the observed butterfly
  diagrams.
  \item When we incorporate  a probably more realistic bipolytropic
  density profile in the tachocline and the radiative zone below the
  convective layer, the results above are maintained. 
\item Finally, if we consider an ellipsoidal tachocline instead of a
  spherical one, we find that when a prolate shape is assumed and the
  magnetic field is stored mostly above the tachocline, then the
  resulting butterfly diagram is in good agrement with the
  observations. 
\item The results above suggest that, although the models with a 
deep meridional flow are, in general, able to produce results 
which are in good agreement with the observations, the global
behavior of the circulation and the  physical mechanism behind it
needs further revision in order to determine where the magnetic
field is really amplified and stor ed (see Guerrero, de Gouveia Dal
Pino \& Munoz 2005, in preparation).
\end{enumerate}

\section{Acknowledgments}
G.A.G. and E.M.G.D.P acknowledge financial support from the
Brazilian Agencies CAPES and CNPq.

\bibliography{biblio}

\end{document}